\documentstyle[11pt]{article}
\newcommand{\newc}{\newcommand}
\newc{\renewc}{\renewcommand}
\newc{\pageyear}[2]{, #1 (19#2)}
\newc{\yearpage}[2]{ (19#2), #1}
\newc{\pandy}[2]{\yearpage{#1}{#2}}
\newc{\ibid}[3]{{\em ibid.\/}\ {\bf #1}\pandy{#2}{#3}}
\newc{\pl}[3]{Phys.\ Lett.\ {\bf #1}\pandy{#2}{#3}}
\newc{\plb}[3]{Phys.\ Lett.\ B\ {\bf #1}\pandy{#2}{#3}}
\newc{\prl}[3]{Phys.\ Rev.\ Lett.\ {\bf #1}\pandy{#2}{#3}}
\newc{\np}[3]{Nucl.\ Phys.\ {\bf #1}\pandy{#2}{#3}}
\newc{\npb}[3]{Nucl.\ Phys.\ {\bf B#1}\pandy{#2}{#3}}
\newc{\rpp}[3]{Rep.\ Prog.\ Phys.\ {\bf #1}\pandy{#2}{#3}}
\newc{\yf}[3]{Yad.\ Fiz.\ {\bf #1}\pandy{#2}{#3}}
\newc{\sjnp}[3]{Sov.\ J.\ Nucl.\ Phys.\ {\bf #1}\pandy{#2}{#3}}
\newc{\zph}[3]{Z.\ Phys.\ {\bf #1}\pandy{#2}{#3}}
\newc{\ijmpa}[3]{Int.\ J.\ Mod.\ Phys.\ A {\bf #1}\pandy{#2}{#3}}
\newc{\prd}[3]{Phys.\ Rev.\ D {\bf #1}\pandy{#2}{#3}}
\newc{\zetp}[3]{Zh.\ Eksp.\ Teor.\ Fiz.\  {\bf #1}\pandy{#2}{#3}}
\newc{\jetp}[3]{Sov.\ Phys.\ JETP {\bf #1}\pandy{#2}{#3}}
\newc{\jmp}[3]{J.\ Math.\ Phys.\  {\bf #1}\pandy{#2}{#3}}
\newc{\bm}[1]{\mbox{\boldmath $#1$}}
\newc{\bvec}{\bm{B}}
\newc{\hvec}{\bm{H}}
\newc{\evec}{\bm{e}_3}
\newc{\ezvec}{\bm{e}_3}
\newc{\exvec}{\bm{e}_1}
\newc{\eyvec}{\bm{e}_2}
\newc{\rhovec}{\bm{\rho}}
\newc{\tauvec}{\bm{\tau}}
\newc{\sigvec}{\bm{\sigma}}
\newc{\Avec}{\bm{A}}
\newc{\avec}{\bm{a}}
\newc{\abvec}{\overline{\bm{A}}}
\newc{\zvec}{\bm{Z}}
\newc{\yvec}{\bm{Y}}
\newc{\xvec}{\bm{X}}
\newc{\nvec}{\bm{n}}
\newc{\svec}{\bm{s}}
\newc{\xivec}{\bm{\xi}}
\newc{\rvec}{\bm{r}}
\newc{\txt}{\textstyle}
\newc{\dsp}{\displaystyle}
\newc{\scr}{\scriptstyle}
\newc{\teeny}{\scriptscriptstyle}
\newc{\be}[1]
{\begin{equation} \mbox{$\label{#1}$}}
\newc{\bea}[1]
{\begin{eqnarray} \mbox{$\label{#1}$}\txt}
\newc{\ee}{\end{equation}}
\newc{\eea}{\end{eqnarray}}
\newc{\non}{\nonumber}
\newc{\newl}{\non\\*}
\newc{\beanon}{\begin{eqnarray*}}
\newc{\eeanon}{\end{eqnarray*}}
\newc{\eref}[1]{Eq.~(\ref{#1})}
\newc{\pref}[1]{(\ref{#1})}
\newc{\ie}{i.\frenchspacing e.\nonfrenchspacing \ }
\newc{\eg}{e.\frenchspacing g.\nonfrenchspacing \ }
\newc{\cc}{\mbox{$c.c.$ }}
\newc{\cf}{cf.\ }
\newc{\comma}{\hspace{.15cm} {\rm ,}\hspace{-.15cm}}
\newc{\period}{\hspace{.15cm} {\rm .}\hspace{-.15cm}}
\newc{\semic}{\hspace{.15cm} {\rm ;}\hspace{-.15cm}}
\newc{\eps}{\epsilon}
\newc{\epsc}{\epsilon^{\ast}}
\newc{\lam}{\lambda}
\newc{\lamc}{\lambda^{\ast}}
\newc{\muc}{\mu^{\ast}}
\newc{\ch}{\raisebox{.5ex}{$\chi$}}
\newc{\rp}{\mbox{$r${\small'}}}
\newc{\half}{{{1}\over{2}}}
\newc{\quarter}{{{1}\over{4}}}
\newc{\third}{{{1}\over{3}}}
\newc{\rec}[1]{{{1}\over{#1}}}
\newc{\rsqrt}[1]{\mbox{$\displaystyle {{1}\over{\sqrt{#1}}}$}}
\newc{\im}{{\rm Im}~}
\newc{\re}{{\rm Re}~}
\newc{\zc}{z^\ast}
\newc{\rtwo}{{\cal R}^2}
\newc{\rthree}{{\cal R}^3}
\def\bigone{\relax{1\kern-.14cm 1}}
\newc{\colvector}[2]{\left(\begin{array}{c} #1\\*#2\end{array}\right)}
\newc{\swap}{\leftrightarrow}
\newc{\gtapprox}
{{\,}^{\raisebox{-.4ex}{${\scriptstyle>}$}}_
{\raisebox{.4ex}{${\scriptstyle\sim}$}}\,}
\newc{\ltapprox}
{{\,}^{\raisebox{-.4ex}{${\scriptstyle<}$}}_
{\raisebox{.4ex}{${\scriptstyle\sim}$}}\,}

\newc{\sign}[1]{\mbox{${\rm sign}(#1)$}}
\newc{\del}{\mbox{$\bm{\nabla}$}}
\newc{\dv}[1]{\mbox{$d^{\,2}\!#1$}}
\newc{\order}[1]{{\cal O}(#1)}
\newc{\dd}[2]{{{\partial #1}\over{\partial #2}}}
\newc{\ddd}[2]{{{{\partial}^2 #1}\over{\partial {#2}^2}}}
\newc{\dddd}[3]{{{{\partial}^2 #1}\over
{\partial #2 \partial #3}}}
\newc{\smalldag}{{\mbox{{\footnotesize\dag}}}}
\newc{\bra}[1]{\langle#1|}
\newc{\ket}[1]{|#1\rangle}
\newc{\ftensor}[3]{\mbox{$\partial_{#2} #1_{#3} -
\partial_{#3} #1_{#2}$}}
\newc{\ev}[1]{\langle #1 \rangle}

\renewc{\thesubsection}{\arabic{subsection}}
\def\zno {{${\rm Z}_{\rm NO}\,$}}

\def\thw{{\theta_{\rm w}}}
\newc{\ewgroup}{\mbox{${\rm SU(2)}\times {\rm U(1)}$}}
\newc{\uone}{\mbox{${\rm U(1)}$}}
\newc{\sdot}{\dot{s}}
\newc{\vdot}{\dot{v}}
\newc{\xdot}{\dot{x}}
\newc{\ydot}{\dot{y}}
\newc{\itwo}{{{i}\over{2}}}

\def\pmb#1{\setbox0=\hbox{#1}%
\kern -.025em\copy0\kern -\wd0
\kern .05em\copy0\kern -\wd0
\kern -.025em\copy0\kern -\wd0\box0}

\def\ab {{\overline \alpha}}
\def\vb {{\overline v}}
\def\yb {{\overline y}}
\def\Yb {{\overline Y}}

\def\nb {{\overline n}}
\def\mb {{\overline m}}
\def\ob {{\overline \omega}}
\def\dr {{d\over dr}}

\renewc{\thefootnote}{\fnsymbol{footnote}}
\begin{document}
\begin{flushright}
NORDITA -- 95/25 P\\
hep-ph/9503460
\end{flushright}
\vspace*{1cm}
\noindent
{\Large\bf Instability of a Nielsen-Olesen Vortex Embedded\vspace*{.7mm}\\
in the Electroweak Theory:\\
II. Electroweak Vortices and Gauge Equivalence\footnote
{Submitted to Mod.\ Phys.\ Lett.\ A as {\it ``Electroweak Vortices and Gauge
Equivalence.''\/}}}

\vspace*{1cm}\noindent
{Samuel W. MacDowell$^1$ and Ola T\"{o}rnkvist$^2$}

\vspace*{5mm}\noindent
$^1$~{\small Yale University, Sloane Physics
Laboratory, New Haven, CT 06520, USA}\\
$^2$~{\small NORDITA, Blegdamsvej 17,
DK-2100 Copenhagen \O, Denmark}

\renewc{\thefootnote}{\arabic{footnote}}
\footnotetext[1]{{\tt macdowell@yalph2.physics.yale.edu}}
\footnotetext[2]{{\tt tornkvist@nbivax.nbi.dk}}
\vspace*{1cm}\noindent
{\bf Abstract}.
Vortex configurations in the electroweak gauge theory are investigated.
Two gauge-inequivalent solutions of the field equations, the Z and  W vortices,
have previously been found. They correspond to embeddings of the abelian
Nielsen-Olesen vortex solution into a U(1) subgroup of \ewgroup.
It is shown here that any electroweak vortex solution can be mapped into
a solution of the same energy with a vanishing upper component
of the Higgs field. The correspondence is a gauge equivalence for
all vortex solutions except those for which the winding numbers of the upper
and lower Higgs components add to zero. This class of solutions, which
includes the W vortex, instead corresponds to a singular solution in the
one-component gauge. The results, combined with numerical investigations,
provide an argument against the existence of other vortex solutions in
the gauge-Higgs sector of the Standard Model.

\vspace*{3ex}
\noindent
{\bf 1. Introduction.}
The electroweak \ewgroup{} gauge theory \cite{WeinSal}
is known to admit at least two distinct vortex solutions, the Z vortex
\cite{NamVac} and the W vortex \cite{Wvort}. For each of the solutions,
a subset of the fields satisfies the field equations of an abelian
Nielsen-Olesen vortex \cite{NieO73}, while the other fields satisfy their
equations trivially.  The Z vortex, for example, is represented by an azimuthal
Z field $Z_\varphi(\rho)$ and a lower component of the Higgs field
$\Phi_2 = \Phi(\rho) \exp(in\varphi)$ which together satisfy the field
equations
of the abelian Higgs model. Here $(\rho,\varphi)$ are polar coordinates of the
position vector $\rhovec$ perpendicular to the vortex, and $n$ is the winding
number.

For the physical value of the Weinberg angle, $\sin^2 \thw\approx 0.23$, the Z
vortex of
winding number $n=1$ has been shown to be unstable with respect to
perturbations in the charged W field \cite{JamPV,Perk93} with angular momentum
$m=-1$, corresponding to the pair production of oppositely charged W bosons
with angular momenta $(m,-m)$. This instability comes about because of the
interaction of the Z field strength with the anomalous magnetic moment
of the W boson.

Following suggestions that the instability of the Z vortex might lead to
a vortex state with a condensate of W-boson pairs \cite{Perk93,TornOle}
similar to the condensate formed in a strong uniform magnetic field
\cite{Magn}, a search for new vortex-like solutions with cylindrical symmetry
was initiated. In one of these searches, undertaken by Ach\'{u}carro et al.\
\cite{AchGHK93}, both components of the Higgs doublet were allowed to vary,
leading to a gauge redundancy.
Numerical solutions including W fields were found for $n>1$ and $m=-1$,
but they were able to show that the solution for each $n$ is gauge equivalent
to a Z vortex with winding number $n-1$.

In a previous paper \cite{MT94} we have investigated both the instability of
the Z vortex and the existence of solutions with W fields in a different gauge,
 defined by the condition that the upper component of the Higgs doublet
vanishes.
In this one-component gauge, we  showed analytically that
the Z vortex with a general winding number $n$
is (in a certain domain of the parameters $\beta,\gamma$) unstable under
W-production in a state of angular momentum $m$ such that $-2n<m<0$. On the
other hand, it was demonstrated that solutions with W fields can exist,
in this gauge, only for $m$ outside this range.

The purpose of the present paper is to establish an energy-preserving
correspondence between vortex solutions in the two-component gauge and those
in the one-component gauge.
It will be shown that any two-component vortex solution with winding number
$n$,
with the exception of $m=-2n$ solutions, is gauge equivalent to a regular
solution in the one-component gauge with winding number either $n$ or $m+n$.
This is a generalization of the result obtained in Ref.\ \cite{AchGHK93}
for $m=-1$. The class of solutions with $m=-2n$, which includes the W vortex,
instead corresponds to solutions in the one-component gauge with the same
energy, but with SU(2) vector potentials behaving as $1/\rho$ near the origin.
The solutions in the two gauges are related by a singular gauge transformation.

Because of this gauge equivalence of solutions the search for new solutions
can be completely carried out in the one-component gauge.
A numerical search for new solutions with W's in this gauge, for
the phase indices $n=1$ and allowed values of $m$ ($m=1,0,-3$), was done
by the authors with negative results. Using both variational methods and
numerical methods for solution of the non-linear differential equations, only
the already known \zno solutions were reproduced. This gives a rather strong
indication that the gauge-Higgs sector of the Electroweak Theory
admits only the already known Z and W vortex solutions.

\vspace*{2ex}\noindent
{\bf 2. Nonabelian Vortex.}
Let $g,g'$ be the coupling constants for the groups SU(2) and U(1)
respectively. They are related to the Weinberg angle $\thw$ and the
electromagnetic charge $e$ by $g\sin\thw=g'\cos\thw =e$. The physical
gauge fields are related to the gauge potentials ${\bf V}^a$ and ${\bf V}'$
associated with the groups SU(2) and U(1) by
$${\bf A}={\bf V'}\cos \thw+{\bf V}^3\sin \thw,\;\;
{\bf Z}={\bf -V'}\sin \thw+{\bf V}^3\cos \thw,\;\; {\rm and}\;\;
{\bf W}=({\bf V}^1-i{\bf V}^2)/\sqrt{2}.$$
Let us define a dimensionless vector
${\bf r} =\rhovec\,g \Phi_0 /({\sqrt{2}}\cos\thw)\equiv
\rhovec M_Z$
with polar coordinates $r,\varphi$,
where $\Phi_0$ is the Higgs vacuum expectation value.

We construct the most general time-independent electroweak vortex ansatz by
letting the circle
at $r=\infty$ map to an arbitrary U(1) subgroup of \ewgroup{} and demanding
that all fields be periodic in the azimuthal angle. The Higgs field is then
given by
\be{higgsdef}
\Phi = \Phi_0 R(r) \exp\left[i \left({{m}\over{2}} \avec\cdot\sigvec + (n +
{{m}\over{2}})\right)\varphi\right] \colvector{0}{1} \simeq
\Phi_0 \colvector{-i s_1(r) e^{i(m+n)\varphi}}{s_2(r) e^{i n\varphi}}\comma
\ee
where $\avec=(\sin\alpha\cos\lambda,\sin\alpha\sin\lambda,\cos\alpha)$ is a
unit vector that may depend on $r$, $\sigvec=(\sigma_1,\sigma_2, \sigma_3)$
is a vector of Pauli matrices, $m,n$ are integers, $s_1=R \sin(\alpha/2)$,
$s_2=R \cos(\alpha/2)$
and $R(\infty)=1$.

In what follows we shall use the expression on the right-hand side which was
obtained
from the general expression on the left by a gauge rotation depending only on
$r$. Cylindrical symmetry of the energy density requires the field ${\bf W}$ to
be of the form
\be{Wdef}
{\bf W}\cos\thw=\Phi_0 [u(r){\bf e}_r+iv(r){\bf
e}_\varphi]\exp(im\varphi)\period
\ee
A change in the relative phase of $u$ and $v$ affects the radial components
$A_r$ and $Z_r$.
It can be shown that one may choose $u,v$ real and $A_r=Z_r=0$ without loss of
generality.
Let us then introduce a set of functions $X,Y,Z$ defined by
\be{XYZdef}
V_{\varphi}^{3}\cos\thw/\sqrt{2}=\Phi_{0}Y(r),\;\;
V_{\varphi}'\sin\thw/\sqrt{2}= \Phi_{0}X(r),\;\;
Z_\varphi/\sqrt{2}=\Phi_0(Y-X)=\Phi_0 Z(r).
\ee
It is also convenient to use a set of auxiliary fields
\be{yxzdef}
y=Y-{\txt{{m}\over{2r}}} ,\quad
x=X-{\txt{{m}\over{2r}}-{{n}\over{r}}},\quad
z=Z+{\txt{{n}\over{r}}}=y-x
\ee
and the parameters $\beta= {(M_H/M_Z)}^2$,
$\gamma={(M_W/M_Z)}^2=\cos^{2}\thw\period$

The energy density in terms of these static fields and the new variables ${\bf
r}$
takes the form
\bea{edens}
{\cal H}
&\!\!\!=\!\!\!&\Phi_0^2{\left\{(\sdot_1+us_2)^2+
(\sdot_2-us_1)^2+((y+x)s_1-vs_2)^2
+((y-x)s_2+vs_1)^2\right.}\non\\*
{}~&+&{\txt\left.{1\over 4}\beta (s_1^2+s_2^2-1)^2
 + {1\over \gamma}{\left(\vdot+{{v}\over{r}}+2yu\right)}^2
+{1\over{\gamma}}\left(\ydot+{{y}\over{r}}
-2uv\right)^2  +{{1}\over{1-\gamma}}{\left(\xdot
+{{x}\over{r}}\right)}^2\right\}}\comma
\eea
where a {\it dot} indicates differentiation with respect to $r$.
In addition to gauge invariances the action in this model is invariant under
charge conjugation. This implies the invariance of the energy density under
the following substitutions:
$$ y\to-y,\quad x\to-x, \quad u\to-u, \quad s_1\to-s_1\quad {\rm and} \quad
(n\to-n,\ m\to-m).$$
Consequently it is sufficient to consider only positive values of $n$.

\vspace*{2ex}\noindent
{\bf 3. Correspondence of Gauges.}
It is easy to verify that the Euler-Lagrange equations for the variational
principle $\delta \int {\cal H}d^2{\bf r}=0$ are not all independent.
The equation for the field $s_1$ is implied by the other equations.
In the case $s_1=0$ it becomes the integrability condition of the other
equations. This means that the gauge, as expected, has not been completely
fixed in our ansatz. The form of the ansatz (and the energy) is in fact
invariant under two gauge transformations $U_0$ or $U_{\pi}$, where
\be{udef}
U_0(\xi) ={\txt
\exp(\itwo\sigma_3 m\varphi)\exp(\itwo\sigma_1\xi)\exp(-\itwo\sigma_3m\varphi);
\quad U_{\pi}(\xi)\equiv i \sigma_1 U_0(\xi-\pi)\period}
\ee
For these transformations to be nonsingular the gauge parameter $\xi(r)$ must
satisfy the boundary conditions $\xi(0)=0$ or $\xi(0)=\pi$ respectively.
Under the transformation $U_{\pi}$ the phase indices $m$ and $n$ change to
$\nb=n+m$ and $\mb=-m$ while for $U_0$, $\nb=n$ and $\mb=m$.
In either transformation, $\alpha\to\ab=\alpha-\xi$,
$Y\to\Yb$, $y \to \yb$ and $v \to \vb$, where
\be{yvtrans}
 \yb\equiv \Yb - {{\mb}\over{2r}} = y \cos\xi + v\sin\xi\ ;
\quad\quad \vb=-y\sin\xi + v\cos\xi\period
\ee
It is then obvious that, whenever allowing a two-component Higgs field, the
gauge freedom can be used to set $\vb(r)=0$. Vice versa, by applying
$U_0(\omega)^{-1}$ or $U_{\pi}(\omega)^{-1}$ to a configuration with $\vb=0$,
$y$ and $v$ may always be parametrized by $y=G\cos\omega$, $v=G\sin\omega$,
where $\omega(0)=0\ {\rm or}\ \pi$ respectively. In this parametrization,
a gauge transformation with parameter $\xi$ corresponds to a shift
$\omega\to\ob=\omega-\xi$.

Before proceeding we shall identify the embedded Nielsen-Olesen vortex
solution corresponding to the Z vortex. This
\zno solution has only the lower component of the Higgs
field $s_2 = f_{{\scr NO}}(r) \exp(in\varphi)$  and fields
$Z=-v_{{\scr NO}}(r)/r$, $X=(\gamma-1)Z$, $Y=\gamma Z$,
and corresponds to $\alpha\equiv\omega\equiv 0$. Here the functions $f_{{\scr
NO}}$
and $v_{{\scr NO}}$ are those
defined in Ref.~\cite{JamPV}.

One finds that in the new variables $R,\alpha,G,\omega$ the action density
\pref{edens}
can
be written as:
\bea{edensvar}
{\cal H}&\!\!\!=\!\!\!&\Phi_0^2{\Big \{}{\dot R}^2+R^2{{\dot \alpha}^2\over 4}
+{1\over{\gamma}}(({\dot G}+{G\over r})^2+G^2{\dot \omega}^2)
+{\beta \over 4}(R^2-1)^2+{1\over {1-\gamma}}({\dot x}+{x\over r})^2 \non\\*
&& +(x^2+G^2)R^2-2R^2Gx\cos(\alpha-\omega)+(R^2 +{4G^2\over{\gamma}})u^2
+u(R^2{\dot \alpha}+{4G^2\over{\gamma}}{\dot \omega}){\Big \}}
\eea
The Euler-Lagrange equation for $u$ gives
\be{uconstr}
(R^2+{4G^2\over \gamma})u+R^2{{\dot \alpha}\over 2}
+{2G^2\over \gamma}{\dot\omega}=0
\ee
Replacing $u$ as given by this equation into the action one obtains, after some
algebraic manipulation, the following constrained action density:
\bea{hc}
{\cal H}^{(c)}&\!\!\!=\!\!\!&\Phi_0^2{\left\{{\dot R}^2+{1\over \gamma}(\dot
G+{G\over r})^2
+{1\over{1-\gamma}}({\dot x}+{x\over r})^2+{\beta \over
4}(R^2-1)^2+(x^2+G^2)R^2\right.}
\non\\*
&&-{\left.2R^2Gx\cos(\alpha-\omega)
+{R^2G^2\over {\gamma R^2+4G^2}}({\dot \alpha}-{\dot \omega})^2 \right\}}
\eea
This expression depends on $\alpha$ and $\omega$ only in the combination
$(\alpha-\omega)$ which is again the evidence of the lack of gauge fixing.

By setting $\alpha\equiv 0$ or $\omega\equiv 0$ one obtains two
distinct gauges, one-component and two-component Higgs gauges, C1$[\omega]$ and
C2$[\alpha]$ respectively. The argument in a square bracket indicates which
field remains dynamical in each gauge.
Now one can see from \eref{hc} that the Euler-Lagrange equations
for the variational principle $\delta \int {\cal H}^{(c)}d^2{\bf r}$ are the
same in the two gauges C1$[\omega]$ and C2$[\alpha]$
provided that one identifies the variable $\omega$ in the first gauge with
$\pm\alpha$ in the second gauge. This will allow us to establish a one-to-one,
energy preserving correspondence between solutions of the field equations
in the two gauges.

In the gauge C1$[\omega]$, if $m\neq 0$, one has to impose the boundary
condition $\omega(0)=(0\ {\rm or}\ \pi)$ in order to exclude a pole of $v$
at $r=0$.
Then the boundary conditions for $\alpha(r)$ in the C2$[\alpha]$ gauge leading
to a solution with non-singular physical fields should be $\alpha(0)=0$ or
$\alpha(0)=\pi$.
In fact, the Euler-Lagrange equation for $\alpha$ in this gauge is:
\be{alphaeq}
\dr{\Big (}{rR^2G^2\over{\gamma R^2+4G^2}}
\dr\alpha{\Big )}-rR^2Gx\sin\alpha=0
\ee
This equation admits the trivial solutions $\alpha(r)\equiv 0$ and
$\alpha(r)\equiv\pi$
which correspond to the \zno solutions with only the lower component Higgs and
index $n$ or the upper component Higgs and index $l=n+m$.
In the second case the solution in the first gauge will be obtained by applying
the gauge transformation $U_\pi(\xi)$ with $\xi=\pi$.
Let us next investigate the possibility of solutions other than those. If they
exist, they would correspond to states with charged W fields.

If $m\ne (0\; {\rm or}\,-2n)$ then near $r=0$, $y=G\sim -m/2r$ and
$x\sim -(2n+m)/2r$. Setting $R=R_0r^p$ and
$\alpha=\alpha_0+\alpha_1 r^q,\;(q>0)$, \eref{alphaeq} reduces to
\be{taylor}
r^{(2p+q-1)}q(2p+q)\alpha_1 - r^{2p-1}m(2n+m)\sin(\alpha_0+\alpha_1r^q)=0
\ee
which, since $q>0$, implies that $\alpha(0)=\alpha_0= (0\; {\rm or}\; \pi)$.

{}From \eref{taylor} we obtain necessary conditions for the existence of
solutions in the C2$[\alpha]$ gauge.

In the case $\alpha(0)=0$, one finds $p=n$ and either $q=m$ or $q=-(2n+m)$.
Since $q$ must be positive it follows that $m$ must be {\it outside\/} the
interval
$-2n<m<0$. This agrees with the condition for existence of solutions in
the C1$[\omega]$ gauge found previously \cite{MT94}. Indeed, solutions
 with $\alpha(0)=0$ would be related to solutions in the C1$[\omega]$ gauge
with $\omega(0)=0$ and the same $n,m$ by the gauge transformation $U_0(\xi)$
where $\xi(r)=\alpha=-\omega$.

In the case $\alpha(0)=\pi$ we obtain $p=|n+m|$ and $q=-|m+n|\pm n >0$. Since
$n$ is positive, it follows that $m$ must now satisfy $-2n<m<0$.
For any allowed $m$ the corresponding solution in the gauge C1$[\omega]$
is obtained by the gauge transformation $U_{\pi}(\xi)$ where
$\xi(r)=\alpha=-\omega$ (modulo $2\pi$). It will have $\omega(0)=\pi$ and phase
indices $\nb=n+m$ and $\mb=-m$.
If one uses charge conjugation invariance to fix the sign of $\nb$ as positive,
$(\nb=|n+m|,\mb=\pm m)$ then the condition $-2n<m<0$ in gauge C2$[\alpha]$ is
equivalent to the condition $\mb$ {\it outside\/} the interval $-2\nb<\mb<0$ in
the
gauge C1$[\omega]$.

If $m=0$ there is no restriction on $\alpha(0)$ (or on $\omega(0)$ in the
C1$[\omega]$ gauge) and the two gauges are equivalent, since then
$U_{0}(\xi)\equiv U_{\pi}(\xi)$ is nonsingular for any value of $\xi(0)$.

If $m=-2n$, then in \eref{taylor} one has to take into account the behavior
$x\simeq x_1r$ of the function $x(r)$ near $r=0$. One finds $q=2$ and
again no restriction on $\alpha(0)$. One can show that in the one-component
gauge C1$[\omega]$, if a solution with charged W fields exists, $v(r)$ must
have a simple
pole and $u(r)$ vanishes at $r=0$. Since the gauge C2$[\alpha]$ was fixed by
the
condition $v(r)=0$ under the assumption that $v(r)$ was regular in any
equivalent gauge it follows that a corresponding solution in the second gauge
can only be related to that in the first by a singular gauge transformation.

If $\gamma=1$, then one must have $x(r)=X(r)=0$ and the gauge group reduces
to SU(2). A solution with $x(r)=0$, corresponding to the embedding of the
Nielsen-Olesen vortex in an SU(2) group, is then expected to exist also for
$\gamma<1$ in the \ewgroup{} gauge theory.
An inspection of the energy density \pref{edens} when $x(r)=0$ shows that
it is invariant under a rotation of the two-component vector $\bm{s}$
defined by $\bm{s}=(s_1,s_2)$. In a two-component gauge one should be able
to use this freedom of rotation to enforce the equation
$x(r)=0$. After obtaining the equation for $x$ from the expression \pref{edens}
for the energy density, one
finds that for $\gamma<1$, a solution $x(r)=0$ implies the condition
$2y(s_1^2-s_2^2)-4vs_1s_2=0$.
A non-singular solution will require $s_1(r)=\pm s_2(r),\;\;v(r)=0$,
that is, it could exist in the C2$[\alpha]$ gauge.
Such a solution, the W vortex, was indeed found in
Ref.~\cite{Wvort}. In the C2$[\alpha]$ gauge and in terms of the fields
$R,G,x,\alpha,\omega$ it is given by:
$$
\omega\equiv 0,\;\;\alpha\equiv\pi/2,\;\; x=0,\;\;
R=f_{{\scr NO},\beta\to \beta/\gamma}(\sqrt{\gamma} r),\;\;
y=G=-v_{{\scr NO},\beta\to \beta/\gamma}(\sqrt{\gamma} r)/r + n/r.
$$
We remark that the W vortex, in this gauge, contains no charged W fields.
The corresponding solution with W's in the C1$[\omega]$ gauge is obtained
by applying
to this solution the singular gauge transformation $U_0({\pi\over 2})$.

The invariance of the energy density under a rotation of the vector
$\bm{s}$ implies that for $\gamma<1$ and for any
$\beta$ the W vortex is a saddle point, since around this extremum
there are directions in function space along which the energy does not change.
This result is in agreement with the conclusion of Klinkhamer and Olesen
\cite{KliOle}.

One concludes from this analysis that except when $n+l=2n+m=0$, every regular
solution of the Euler-Lagrange equations in a two-component gauge, with given
values of the indices ($n,m$) is gauge equivalent to a solution in the
one-component gauge with either indices ($\nb=n,\mb=m$) or ($\nb=n+m,\mb=-m$).
This generalizises a result of Ach\'{u}carro et al.\
\cite{AchGHK93} who found
numerically  a two-component solution with charged W fields for the special
case
$m=-1$ and showed that it was gauge equivalent to the \zno solution with index
$n-1$.

In the exceptional case $2n+m=0$, a solution with charged W's
in the gauge C1$[\omega]$ is
always singular and can only be related to a regular solution in the
C2$[\alpha]$ gauge by a singular gauge transformation.

Regular solutions with charged W's in the one-component gauge must have $m$
outside
the interval $-2n\le m<0$. We have made a rather thorough numerical search for
solutions with W's in this gauge for $n=1$ and $m=1,0,-3$ with negative
results.
This seems to indicate that the Z vortex and the W vortex already found
are the only possible vortex solutions in the gauge-Higgs sector of the
\ewgroup{} Electroweak Theory.

One of us, O.T., would like to thank Poul Olesen for helpful discussions.


\begin{thebibliography}{99}
\setlength{\parsep}{-1mm}
\setlength{\itemsep}{0pt}
\bibitem{WeinSal}
S. Weinberg, \prl{19}{1264}{67};
A. Salam in {\em Elementary Particle Theory\/},
edited by N. Svartholm (Almqvists F\"{o}rlag AB, Stockholm, 1968) p. 367.
\bibitem{NamVac}
Y. Nambu, \npb{130}{505}{77};
T. Vachaspati, \prl{68}{1977}{92}; \ibid{69}{216}{92}.
\bibitem{Wvort}
T. Vachaspati and M. Barriola, \prl{69}{1867}{92};
M. Barriola, T. Vachaspati, M. Bucher, \prd{50}{2819}{94}. In the first
reference,
the W vortex is referred to as a $\tau^1$ string.
\bibitem{NieO73}
H. B. Nielsen and P. Olesen, \npb{61}{45}{73}.
\bibitem{JamPV}
M. James, L. Perivolaropoulos, and T. Vachaspati, \prd{46}{R5232}{92};\
\npb{395}{534}{93}.
\bibitem{Perk93}
W. B. Perkins, \prd{47}{R5224}{93}.
\bibitem{TornOle}
P. Olesen, Niels Bohr Institute preprint NBI-HE-93-58
(Oct. 1993) HEP-PH/9310275; O. T\"ornkvist, Yale Preprint
YCTP-P11-92 (Apr. 1992), HEP-PH/9204235.
\bibitem{Magn}
J. Ambj{\o}rn and P. Olesen, \npb{315}{606}{89};\
\npb{330}{193}{90};\
\ijmpa{5}{4525}{90};
S. W. MacDowell and O. T\"{o}rnkvist,
\prd{45}{3833}{92};
Nils Ola T\"ornkvist, Yale University Ph. D. Thesis (Nov. 1993) RX-1493,
Microfiche UMI-94-15879-mc.
\bibitem{AchGHK93}
A. Ach\'{u}carro, R. Gregory, J. A. Harvey, and K. Kuijken,
\prl{72}{3646}{94}.
\bibitem{MT94}
S. W. MacDowell and O. T\"ornkvist, in Proceedings of the G\"ursey Memorial
Conference I: On
Strings and Symmetries, \.{I}stanbul, Turkey, 6-10 June 1994, ed. M.
Serdaro\u{g}lu,
Springer-Verlag (1995), NORDITA preprint 94/52P, HEP-PH/9410276.
\bibitem{KliOle}
F. R. Klinkhamer and P. Olesen, \npb{422}{227}{94}.
\end{thebibliography}
\end{document}